\documentclass[12pt]{article}
\let\section=\subsection     \let\subsection=\subsubsection                %%
\usepackage{graphicx}

\begin{document}
\begin{center}
   {\large \bf Baryon-baryon interactions in free space}\\[2mm]
   {\large \bf and baryonic medium}\\[5mm]
   C.~Keil and H.~Lenske\\[5mm]
   {\small \it  Institut f\"ur Theoretische Physik I, Univ. Gie\ss en \\
   Heinrich-Buff-Ring 16, D-35392 Gie\ss en, Germany \\[8mm] }
\end{center}

\begin{abstract}\noindent
  We study the interactions of the octet baryons in a relativistic
  meson exchange approach connecting free-space and in-medium
  interactions with those in finite (hyper)nuclei. A
  short sketch of the model including the relativistic T- and G-matrix
  and the mapping of the G-matrix onto density dependent
  vertex functionals in the density dependent relativistic hadron
  field theory (DDRH) is given. From new spectroscopic data on medium mass
  hypernuclei we extract the single particle spectrum including
  spin-orbit splitting. An extraction of the $\Lambda\omega$
  vector- and tensor vertex as well as the $\Lambda\sigma$-vertex
  shows a strong breaking of SU(3) symmetry.
\end{abstract}

\section{Introduction}
The presented approach aims at a microscopic link between interactions
for free baryon-baryon scattering, their in-medium interaction in
infinite matter and in finite nuclei. Such a comprehensive description
is essential for extrapolations to not yet experimentally observed or
even never observable regions of the (hyper)nuclear chart, infinite
hypermatter in neutron stars and baryon-baryon interactions in heavy
ion collisions. In the nuclear sector there is abundant information
available, i.e., the coupling constant can be fixed from free NN
scattering and then propagated to finite nuclei where calculations of
their properties and comparison to experimental data yield an important check
of consistency. In the hyperon sector the model can
be used then to extract from single particle spectra of
hypernuclei free hyperon coupling constants which are unaccessible otherwise.

\section{The density dependent relativistic hadron field theory}
Our approach is based on the covariant meson exchange framework. To
generate the effective microscopic interaction in finite nuclei an approach
using T-matrix theory with a three dimensionally reduced two-nucleon propagator
is applied. In this way the same mesons with the same bare couplings are
used throughout the model. The in-medium interaction is calculated in
Dirac-Brueckner theory, i.e., a T-matrix calculation in infinite
matter where the intermediate two-baryon propagators are Pauli-blocked
and dressed by Hartree-Fock self-energies. Since a G-matrix
calculation for finite nuclei is technically not feasible DDRH theory
is used to apply the G-matrix interactions in calculations of finite
nuclei.

DDRH theory \cite{Fuchs:1995as} is a Walecka type model designed for
relativistic mean-field (RMF) calculations of finite nuclei using G-matrix
interactions. Contrary to the conventional RMF models density dependent vertex
functionals $\Gamma(\hat\rho)$ are used instead of constant couplings and
meson self-interactions to account for finite density effects. $\hat\rho$ is a
bilinear of the baryonic field operators. Using the self-energies, in-medium
effects of the G-matrix interaction are mapped onto the vertex functionals
\cite{Fuchs:1995as,Hofmann:2000vz}. Deriving 
the Dirac equation from the DDRH Lagrangian yields additional {\it
  rearrangement self-energies}, coming from the variation of the vertex
functionals with respect to the baryonic field operators and describing static
polarizations of the nuclear medium \cite{Fuchs:1995as}. They do not affect
the bulk properties of a nucleus but modify the single particle structure
only. This treatment of the density dependence leads to a covariant field
theoretically and thermodynamically consistent theory (see \cite{Fuchs:1995as}
for the proofs). DDRH theory was applied successfully to magic and exotic
nuclei in RMF approximation \cite{Fuchs:1995as,Hofmann:2000vz}.

\section{Hypernuclear structure}
Hypernuclear spectroscopy has made important progress in accuracy in the
recent years, allowing now to extract information about the interaction
systematics between hyperons and nucleons within a microscopic model. With
this understanding of the interaction and upcoming even more precise
$\gamma$-spectroscopy experiments on hypernuclei at KEK, JLab, GSI and MAMI-C
it will even be possible to study more subtle interaction effects beyond the
mean-field which are not accessible in such a clean way using nucleon
spectroscopy and conventional nuclear structure. 

\subsection{DDRH for hypernuclei}
DDRH theory has been extended from the use in pure isospin nuclei to
hypernuclei involving the octet baryons \cite{Keil:1999hk} and to hyperon star
matter \cite{Hofmann:2000mc}. Lacking a G-matrix calculation for the complete
baryon  octet -- work is in progress on that -- an approximate treatment of
the hyperon-meson vertices was obtained from a diagramatic analysis of the
Dirac-Brueckner equations. It was shown \cite{Keil:1999hk} that the functional
shape of the vertices should be approximately the same as that of the nuclear
ones, depending on the density of the respective hyperon iso-multiplet instead
of the nuclear density. The strength has to be rescaled by a function
$R_{Y\alpha}$ which turns out to be approximately constant and can in first
order be related to the free coupling constants as $R_{Y\alpha} =
\frac{g_{Y\alpha}}{g_{N\alpha}}$. 

Single $\Lambda$ hypernuclei are the only well known hypernuclear systems. The
$\Lambda$ being an isoscalar and electrically neutral baryon just couples to
the $\sigma$ and the $\omega$ meson simplifies its theoretical description
tremendously. The two corresponding scaling factors relate directly to the
free couplings $g_{\Lambda\sigma}$ and $g_{\Lambda\omega}$ by the approximate
relations given above. Using spectral information of hypernuclear spectroscopy
including resolved spin-orbit doublets these two constants can be extracted
unambiguously from DDRH RMF calculations for the respective hypernucleus -- or
incompatibilities of the calculated and measured spectral structures will give
definite hints to deformations and dynamics beyond mean field.

\subsection{Evaluation of experimental data}

\begin{figure}
  \begin{minipage}{0.58\linewidth}
    \begin{center}
      \includegraphics[width=8cm,angle=0]{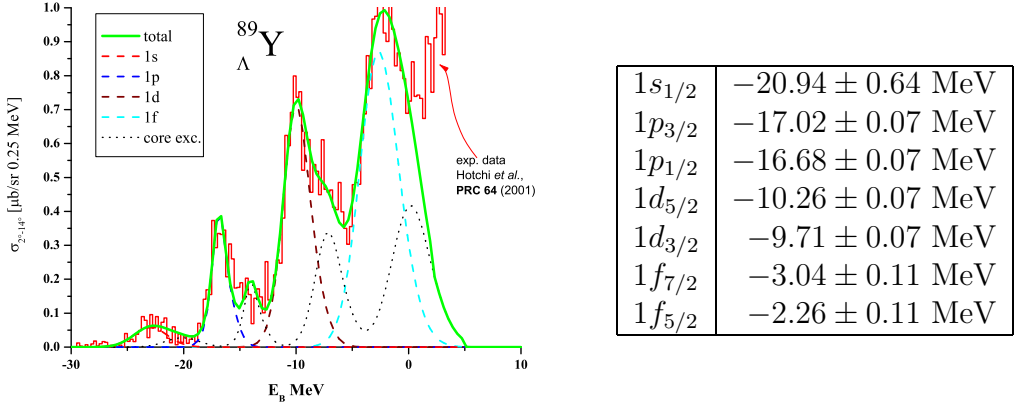}\\
    \end{center}
  \end{minipage}\hfill
  \begin{minipage}{.38\linewidth}
    \begin{tabular}{|l|r|}
      \hline
      $1s_{1/2}$ & $-20.94\pm 0.64$ MeV \\
      $1p_{3/2}$ & $-17.02\pm 0.07$ MeV \\
      $1p_{1/2}$ & $-16.68\pm 0.07$ MeV \\
      $1d_{5/2}$ & $-10.26\pm 0.07$ MeV \\
      $1d_{3/2}$ & $ -9.71\pm 0.07$ MeV \\
      $1f_{7/2}$ & $ -3.04\pm 0.11$ MeV \\
      $1f_{5/2}$ & $ -2.26\pm 0.11$ MeV \\ \hline
    \end{tabular}
  \end{minipage}
\caption{\label{tab:sp-spec}The figure shows the refined fit to extract the
      single particle spectrum from the experimental data, taking also into
      account $jj$ interactions. In the table the extracted single particle
      spectrum of $^{89}_\Lambda Y$ is shown.}
\end{figure}

Recent experimental data measured at KEK on $^{89}_\Lambda Y$ and
$^{51}_\Lambda V$ \cite{Hotchi:2001rx} show a significant broadening of the
single particle peaks as compared to the experimental resolution. The simple
analysis with a two uncorrelated gaussian fit of each single particle peak -
motivated by a spin-orbit (s.o.) splitting assumption - which was performed in
\cite{Hotchi:2001rx} lead to a single particle spectrum which could not be
reproduced by DDRH RMF calculations. Taking into account also the high ground
state spin of the $^{88}Y$ and $^{50}V$ cores with $I^\pi = 4^+, 6^+$,
respectively, we performed a refit of the data. The broad single particle
peaks were modeled by two gaussians of which the individual strength was
constrained by the relative multiplicities of their s.o. states. The width of
each gaussian and the relative distance of their means were fixed by assuming
the spin dependent part of the single particle energies to be $E_s =
E_{ls}\left<l\cdot s\right> + E_{jj}\left<I\cdot j\right>$ ($I$ being the core
spin and $j$ the total angular momentum of the $\Lambda$ state). The s.o. and
$jj$ matrix elements are the same for the whole fitting procedure. In contrast
to the fit with uncorrelated gaussians our fit gets significantly more
stable. A strong reduction of the s.o. strength compared to the simpler fit is
observed being consistent with high precision measurements of the
s.o. splitting in $^{13}_\Lambda C$ \cite{Kohri:2001nc}. The s.o. and $jj$
matrix elements are obtained as $E_{ls}=223\pm 153 \;keV$ and $E_{jj}=61\pm 3
\;keV$.

\subsection{Hypernuclear structure calculation with DDRH}
\begin{figure}
  \begin{center}
    \includegraphics[width=7cm]{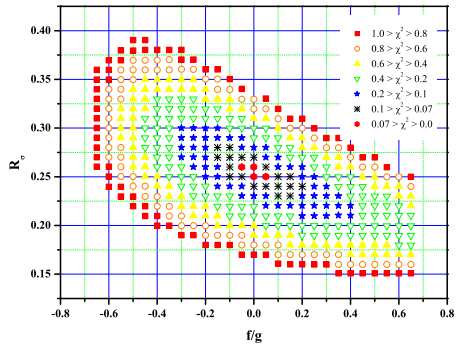}\\
  \end{center}
\caption{\label{fig:chi-sq-plane}$\chi^2$ plane of fitting the $\Lambda$
  meson coupling constants to match the experimental single particle spectrum.}
\end{figure}

The fairly small s.o. splitting brings into the game the formerly already
discussed $\Lambda\omega$ tensor interaction (see, e.g.,
\cite{Mares:1993xb}). This tensor interaction introduces terms proportional to
$\overline{\psi}\sigma^{_\mu\nu}\psi \partial_\mu\omega_\nu$
in the interaction Lagrangian being of the same Dirac structure as the
s.o. interaction in the Dirac equation. According to spin-flavor SU(6) the
tensor vertex "$f$" is minus that of the vector interaction "$g$", leading to
an almost vanishing s.o. splitting.

\begin{figure}
  \begin{center}
    \includegraphics[width=9.5cm]{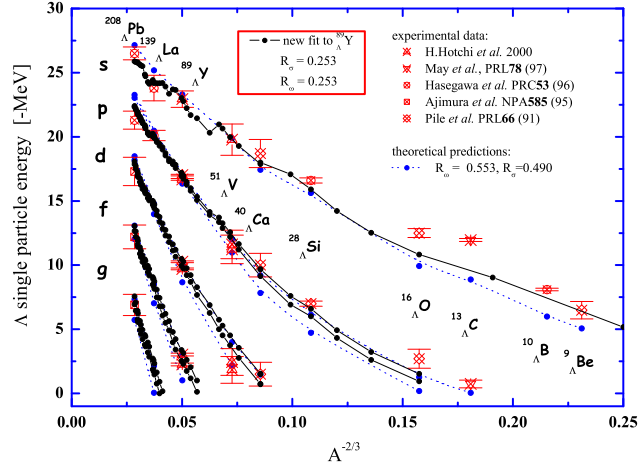}\\
  \end{center}
\caption{\label{fig:spspec}DDRH $\Lambda$ single particle spectra
  compared to the world data.}
\end{figure}

Including the tensor interaction we performed a least squares fit of
calculated DDRH RMF $\Lambda$ single particle energies to the experimental
ones with respect to the variables $R_\sigma$, $R_\omega$ and $f/g$ (which
should be $2/3$, $2/3$ and $-1$, respectively, according to SU(3)). A very
strong linear correlation between $R_\sigma$ and $R_\omega$ is found, induced
by the spectral gross structure (known already from earlier hypernuclear RMF
calculations not considering s.o. splitting \cite{Keil:1999hk}). The favored
region with $\chi^2<1$ is shown in fig.~\ref{fig:chi-sq-plane}. It
is centered around $R_\sigma$(=$R_\omega$) = 0.25 and $f/g$ = 0. This
indicates a substantial violation of SU(3) symmetry. Effects which are still
under investigation and might lead to minor changes are for example the
influence of deformations of the nuclear core. Calculations for the world data
set on hypernuclei with an even-even core is shown in fig.~\ref{fig:spspec}.
The calculation with our new parametrization (dots linked by the solid line)
shows clear improvements especially in the low mass region compared to our
previous parameters. 

Concerning model building the strong violation of SU(3) symmetry is bothering
since SU(3) relations used in the derivation of, e.g., hyperon-hyperon
interactions could no longer be applied.
From the nuclear structure point of view the very weak $\Lambda$ couplings are
in contrast very encouraging. Since the strange sector seems to almost
decouple from the isospin sector it might be possible in future high accuracy
spectroscopic experiments to measure nuclear bulk properties like mass radii
and deformations through $\Lambda$ hypernuclear spectroscopy. 

\section{Conclusions}
The accuracy of the discussed experimental data is already at the limits of
what can be expected from meson spectroscopy of hypernuclei. Still, looking at
the large uncertainties in the determination of coupling constants from these
data, see fig.~\ref{fig:chi-sq-plane}, more precise measurements are
needed. These will be provided by the beginning new era of $\gamma$
spectroscopy experiments on hypernuclei. On the theory side it will be
essential to replace the approximate treatment of the in-medium $\Lambda$
interactions by a fully microscopic one. Work on such a Dirac-Brueckner
G-matrix calculation is in progress.

\section*{Acknowledgments}
This work was supported by the European Graduate School Gie\ss en -Copenhagen, 
DFG contract Le 439/5 and GSI.

\end{document}